\documentclass{ws-rv9x6}
\usepackage{subfigure}     
\usepackage{ws-rv-van}     

\newcommand{\beq}{\begin{eqnarray}}
\newcommand{\eeq}{\end{eqnarray}}
\def\bea{\begin{array}}
\def\eea{\end{array}}

\makeindex
\begin{document}

\chapter[String Cosmology]{String Cosmology - Large-Field Inflation in String Theory}\label{ra_ch1}

\author[A. Westphal]{Alexander Westphal}

\address{Deutsches Elektronen-Synchrotron DESY, Theory Group, \\
D-22603 Hamburg, Germany, \\
alexander.westphal@desy.de}

\begin{abstract}
This is a short review of string cosmology. We wish to connect string-scale physics as closely as possible to observables accessible to current or near-future experiments. Our possible best hope to do so is a description of inflation in string theory. The energy scale of inflation can be as high as that of Grand Unification (GUT). If this is the case, this is the closest we can possibly get in energy scales to string-scale physics. Hence, GUT-scale inflation may be our best candidate phenomenon to preserve traces of string-scale dynamics. Our chance to look for such traces is the primordial gravitational wave, or tensor mode signal produced during inflation. For GUT-scale inflation this is strong enough to be potentially visible as a B-mode polarization of the cosmic microwave background (CMB). Moreover, a GUT-scale inflation model has a trans-Planckian excursion of the inflaton scalar field during the observable amount of inflation. Such large-field models of inflation have a clear need for symmetry protection against quantum corrections. This makes them ideal candidates for a description in a candidate fundamental theory like string theory. At the same time the need of large-field inflation models for UV completion makes them particularly susceptible to preserve imprints of their string-scale dynamics in the inflationary observables, the spectral index $n_s$ and the fractional tensor mode power $r$. Hence, we will focus this review on axion monodromy inflation as a mechanism of large-field inflation in string theory.
\end{abstract}
\body

\section{Introduction}\label{ra_sec1}

Cosmology has long been considered as a speculative field. Envisioning now a combination of a candidate theory of quantum gravity and unification yet to be fully understood, string theory, with early universe cosmology may at first warrant even more caution. However, before rushing to this conclusion we shall start our discussion employing judicious use of the structure of effective field theory as a bottom-up tool and controlled approximations or constructions in string theory. This will lead us to realize that certain phenomena of early and late-time cosmology can be reasonably well embedded and described by certain classes of solutions of string theory.

We will start making the following observation.  String theory is a candidate theory for a fundamental unification of quantum mechanics with general relativity. String theory solutions are typically subject to a requirement of 10 dimensions of space-time. To make contact with our four dimensional large-scale space-time, we need to compactify the six extra-dimensions on an undetectably small internal manifold. However, at energies small compared to the inverse string and compactification length scales string theory reduces to an effective quantum field theory (EFT). This EFT contains gauge fields coupled to fermionic matter and a number of scalar fields as well as general relativity extended by a series of higher-order curvature corrections from the string $\alpha'$-expansion. 

Hence, below string and compactification scales the usual decoupling theorem of QFT ensures the absence of strong string-theoretic effects from, e.g., weak-scale physics -- except if weakly-broken symmetries protect a hierarchical suppression of scales for some of the effects of string theory. Examples for the latter are low-scale broken supersymmetry, and axionic shift symmetries originating from higher-dimensional gauge symmetries of string theory which usually enjoy exponentially suppressed scales of symmetry breaking due to its non-perturbative nature. Without the presence of symmetries surviving unbroken to low scales, we conclude that the only phenomena in cosmology which may warrant or even necessitate a description in a theory of quantum gravity must either live by themselves close to the string scale or display explicit ultra-violet (UV) sensitivity to UV-divergent quantum corrections.

From this argument we discern three such phenomena in need of a description beyond EFT coupled to general relativity:
\begin{itemize}
\item The initial cosmological singularity.\footnote{Note, that even in the presence of eternal inflation space-time is geodesically incomplete to the past, so even then there was a cosmological singularity in our past~\cite{Borde:2001nh}.}
\item A very early period of cosmological inflation. As discussed further below, there is increasingly strong evidence that the CMB temperature fluctuations are down to inflationary scalar quantum fluctuations with almost scale-invariant 2-point function power spectrum $\Delta_{\cal R}^2\sim H^2/\epsilon$. Barring suppression of the first slow-roll parameter $\epsilon$ by tuning, we expect $\epsilon\lesssim 1/N_\star\sim 0.01$ where $N_\star\simeq 60$ denotes the observable span of slow-roll inflation. Hence, from $\Delta_{\cal R}^2\sim 10^{-9}$ we expect slow-roll inflation to happen near the GUT scale.  Moreover, we will see below that slow-roll inflation is inherently UV-sensitive to quantum corrections of the effective potential. This sensitivity is only enhanced by the closeness of the inflationary energy scales to the string scale.
\item The late-time form of dark energy driving the observed contemporary accelerated expansion of our universe. This phenomenon has two aspects of needing a UV completion in some theory of quantum gravity. Firstly, the contribution to dark energy in form of a cosmological constant from quantum field theory is manifestly power law UV-divergent. As such, EFT coupled to general relativity can only explain the present-day amount dark energy $\sim 10^{-122}M_{\rm P}^4$ by manifest and insane amounts of fine-tuning. Hence the need for UV completion. Secondly, a pure cosmological constant generates eternal de Sitter (dS) space-time in general relativity. QFT on eternal dS space-time is not well defined as the S-matrix does not exist. Hence de Sitter space is in need of a more fundamental description, presumably a UV complete theory of quantum gravity.
\end{itemize}

While we do have in recent years some important progress in the description of cosmological singularities in string theory, it is probably still fair to say that a full controlled description of space-time near a realistic cosmological singularity (i.e. the one in the past of null-energy condition satisfying FRW space-times) remains an open question (for reviews see e.g.~\cite{Cornalba:2003kd,Craps:2006yb,Berkooz:2007nm}).

Given the stage set by this general discussion, it is important to recall that it was the advent of several waves of observational progress during the last two decades which have required string theory to confront de Sitter space and the description of cosmological inflation. Moreover, the accuracy of these new experimental results fundamentally changed the speculative status of cosmology, leading to an era of precision cosmology.

At first, highly reliable and increasingly precise measurements of the Hubble Space Telescope (HST) yielded a determination of present-day Hubble parameter $H_0=73.8\pm 2.4 \, \text{km s}^{-1}\text{Mpc}^{-1}$~\cite{Freedman:2000cf,Riess:2011yx}. This milestone enabled the construction of an ultra deep-space distance ladder by type IA supernovae. Systematic observations of such type IA supernovae at high red-shift culminated in the detection of a form of dark energy consistent with extremely small ($\sim 10^{-122}M_{\rm P}^4$) and positive cosmological constant, which drives a late-time accelerated expansion of our Universe~\cite{Riess:1998cb,Perlmutter:1998np}.

In the next step, the space-based satellite missions WMAP~\cite{Hinshaw:2012fq} and PLANCK~\cite{Ade:2013zuv,Ade:2013uln} as well as the ground-based telescopes ACT~\cite{Sievers:2013wk} and SPT~\cite{Story:2012wx,Hou:2012xq} probed the cosmic microwave background (CMB) radiation with unprecedented precision and resolution. Their combined results together with the HST and type IA supernova data led to the concordance model of observational cosmology. In its `essence' this new standard model is consistent with certain simple features of our observed Universe. We find ourselves to a good approximation (sub-\%-level) in a FLRW universe which is spatially flat and undergoes accelerated late-time expansion driven by dark energy. Moreover, the recent CMB missions provided measurements of the two-point function power-spectrum  of the $10^{-5}$-level thermal fluctuations with unprecedented precision. These results now bear increasingly strong evidence that the very early Universe went through a much earlier and extremely rapid phase of accelerated expansion driven by the vacuum energy of a slowly-rolling scalar field, called inflation~\cite{GUTH,STAROBINSKY,LINDE,STEINHARDT} (see e.g.~\cite{Baumann:2009ds} for a recent review).

The recent high-precision data from PLANCK~\cite{Ade:2013zuv,Ade:2013uln} as well as the ground-based telescopes ACT~\cite{Sievers:2013wk} and SPT~\cite{Story:2012wx,Hou:2012xq}, and in particularly the strong limits on the presence of non-Gaussianity in form of a non-vanishing three-point function from the PLANCK mission~\cite{Ade:2013ydc}, are consistent with a picture of simple slow-roll inflation driven by the scalar potential of a single canonically normalized scalar field.

Finally, the recent report from the BICEP2 telescope~\cite{BICEP2} announced the detection of degree angular scale B-mode polarization in the CMB. B-mode polarization in the CMB at such large angular scales can have a primordial origin only from inflationary gravitational waves, so-called tensor modes. However, polarized emission from galactic dust may provide the same type polarization pattern as a foreground contamination~\cite{DUST1,DUST2}. Discrimination may be possible due to different angular power spectra and radio frequency spectra of such dust induced B-modes versus the ones from primordial tensor modes. If the B-mode detection by BICEP2 turns out to be of at least partial primordial original, then the inflationary energy scale would be at the GUT scale, and inflation would have to proceed as a large-field model with trans-Planckian field excursion.\footnote{A very recent analysis of the 2D genus topology statistics and the cross-correlation between the BICEP2 150 GHz data and PLANCK polarization data in the BICEP2 region at 353 GHz~\cite{Planck353Map} by Colley and Gott III allowed for a first more dust-model independent estimate of the dust emission fraction in the BICEP2 signal. This analysis yields the dust part to be a bit less than 50\% and results in $r=0.11\pm 0.04$ ($1-\sigma$) providing $2.5-\sigma$ evidence for $r>0$.\cite{DUST3}}Both features would spell a direct need for embedding inflation into a fundamental candidate theory of quantum gravity such as string theory as we will discuss below.

This progress of the last 15 years forced candidate theories for a fundamental unification of quantum mechanics with general relativity, such as string theory, to accommodate both an extremely tiny positive cosmological constant and the dynamics of slow-roll inflation for the very early Universe. In particular, string theory solutions are typically subject to a requirement of 10 dimensions of space-time. To make contact with our four dimensional large-scale space-time, we need to compactify the six extra-dimensions on an undetectably small internal manifold. This process of compactification produces a huge set of possible suitable manifolds, each of which is accompanied by set of massless 4D scalar moduli fields describing the allowed deformation modes of each manifold. Crucial progress towards describing late-time and early-time (quasi) de Sitter (dS) stages of positive vacuum energy in string theory involved the construction of string vacua with complete stabilization of all geometrical moduli fields and controllable supersymmetry breaking, largely in type IIB string theory (see e.g.~\cite{Grana:2005jc,Douglas:2006es,MODSTABreview} for reviews), and more recently also in heterotic string theory~\cite{Cicoli:2013rwa}. 

Hence, the outline of our discussions proceeds as follows. We will start with a discussion of the basic ingredients of inflation and string theory in sections~\ref{sec_inf} and \ref{sec_strings}.

Section~\ref{sec_stringinf} will discuss inflation in string theory, emphasizing the necessity of having controlled string vacua with full moduli stabilization before even beginning to search for inflationary regimes. Some of the major progress in recent years involved the construction~\cite{KKLT,LENNY03,LVS,KAHLERUPLIFT} of controlled type II string flux compactifications (see e.g.~\cite{Grana:2005jc,Douglas:2006es} for a review) with full moduli stabilization and positive vacuum energy necessary to make contact with cosmologically viable 4D space-time descriptions incorporating  the observed late-time acceleration. A typical compactification has ${\cal O}(100)$ moduli fields allowing for an exponentially large number of possible combinations of fluxes. This led to the discovery of an exponentially large landscape of isolated dS vacua in string theory~\cite{LENNY03}.

Within the extant constructions of string theory dS vacua inflation must then arise from a flat region of the moduli scalar potential supporting the criteria of slow-roll by a flat potential, or higher-derivative kinetic terms originating from mobile D-branes. Since we wish to focus on connecting string-scale physics to the observables of inflation as closely as possible, this motivates us to concentrate our discussion on large-field inflation in the form of axion monodromy~\cite{MSW,KALOPER08} or aligned natural inflation~\cite{KNP,Kappl:2014lra}. The book by Baumann and McAllister~\cite{STRINGINF}contains an exhaustive discussion of the many models of small-field inflation in string theory (e.g. warped D3-brane inflation, D3-D7 inflation, K\"ahler moduli inflation, fibre inflation, DBI inflation, etc.) following the seminal work of Kallosh et al.~\cite{KKLMMT}, which we consequently omit here.

Our discussion will need to include corrections to a scalar potential driving slow-roll inflation. They will arise generically due to dimension-six operators from radiative corrections or integrating out massive states, rendering inflation UV sensitive. Hence, candidate fundamental theories of quantum gravity such as string theory are necessary for a full description of inflation beyond the limits of effective field theory.

Using this stage, we will try explain the general mechanism and structure of axion monodromy which underlies the various models of large-field inflation with axions arising from higher-dimensional gauge-fields in string theory (see e.g.~\cite{Baumann:2009ni,STRINGINF} for very recent reviews). I will openly say at this point, that you will find the most probably definitive reference for inflation in string theory in the recent book by Baumann and McAllister~\cite{STRINGINF}. 


\section{Basics of Inflation}\label{sec_inf}

Inflation is a period of quasi-exponential expansion of the very early Universe, invented originally by Guth~\cite{GUTH} to overcome several initial condition problems of the conventional hot big bang cosmology. A short discussion of the two most pertinent of these, the horizon and flatness problems, will serve us well to illuminate the essence of and need for inflation.

The horizon problem of the hot big bang has its root in the differing behavior of the causal horizon distance and the stretching of physical length scales in an expanding universe driven by radiation or matter. 

\subsection{Classical theory}\label{sec_infclass}

To begin, we recall a few basic facts about the description of spatially homogeneous and isotropic expanding space-times in General Relativity. In an FRW universe driven by any type of matter or energy in perfect fluid form except vacuum energy, we have the energy density diluting as $\rho\sim a^{-3 (1+w)}$. Here $w$ denotes the equation of state of the fluid dominating the energy budget. The 1st Friedmann equation governing an expanding FRW universe, possibly with spatial curvature denoted by a parameter $k=0,\pm 1$ reads
\beq
H^2\equiv \left(\frac{\dot a}{a}\right)^2=\frac13\rho-\frac{k}{a^2}\quad.
\eeq
In a spatially flat universe $k=0$ this describes an expansion following
\beq
a\sim t^p\quad,\quad p=\frac{2}{3(1+w)}\quad{\rm and}\quad H=\frac{p}{t}\quad.
\eeq
At $t=0$ we have $a=0$, implying a curvature singularity, that is, the initial big bang singularity. The exception to the above is pure vacuum energy, e.g. the potential energy of a scalar field, which has $w=-1$ and thus $H=const.$ and $a\sim e^{Ht}$ (i.e. formally $p\to\infty$).

A given physical length scale, e.g. radiation of a given wavelength $\lambda_{ph}$, will stretch with the expansion of an FRW universe as
\beq
\lambda_{ph}(t)=\frac{\lambda_{ph}(t_0)}{a(t_0)}\,a(t)\quad.
\eeq
Hence, we can think about the continually stretching physical length scales in terms of associated 'comoving' length scales $\lambda\equiv \lambda_{ph}/a$, that is with reference length scales obtained by scaling out the 'stretching' with the scale factor $a$.

We now compare a comoving length scale with the comoving light-travel distance since the initial singularity, the comoving horizon distance $\tau$
\beq
\tau=\int \frac{dt}{a}=\int\frac{d\ln a}{aH}\quad.
\eeq
We see from the preceding discussion that for any fluid with $w>-1$ driving the expansion we get
\beq
(aH)^{-1}\sim \frac1p a^{\frac{1-p}{p}}\quad\Rightarrow\quad \tau=\frac{1}{1-p}a^{\frac{1-p}{p}}\sim \frac{1}{aH}
\eeq
while for $w=-1$ directly $\tau=1/aH$.

The last result displays the horizon problem: An expanding universe driven by a fluid with $w> -1/3$ or equivalently $p<1$ has a  \emph{growing} comoving horizon $(aH)^{-1}\sim a^{(1-p)/p}$ with $(1-p)/p>0$ for $p<1$. All comoving scales $\lambda$ enter the horizon from outside from the past towards the future. Therefore, all observable length scales were outside the horizon and thus out of causal contact at sufficiently early times. For the CMB sky this corresponds to patches separated by more than about 1 degree. Yet these patches are all at the same temperature to better than 1 part in $10^4$. Why?

The problem arises from the integral in $\tau$ and the fact that $1/aH$ is growing with $a$ for $w>-1/3$ or $p<1$. The integral gets its main contribution from late times. The problem was solved if at an earlier time we can arrange for the comoving horizon $1/aH$ to \emph{decrease} with increasing $a$. Comoving length scales would then \emph{leave} the horizon at an early time, and then re-enter later on, after the expansion has changed from decreasing comoving horizon to the increasing comoving horizon of matter or radiation domination. Therefore, at a very early time all observable comoving length scales would have been inside the horizon and in causal contact, despite leaving the horizon later on for quite a while.

Since in a expanding universe we have $\dot a>0$ always, this means we need arrange for an early period where
\beq
\frac{d}{dt}\left(\frac{1}{aH}\right)<0\quad{\rm for}\quad t<t_e
\eeq
and positive thereafter. We denote with $t_e$ the time when this early phase of a decreasing comoving horizon ended.

What does this early phase mean? We evaluate $\frac{d}{dt}(aH)^{-1}$ and by differentiating in turn the Friedmann equation we arrive at
\beq
\frac{d}{dt}\left(\frac{1}{aH}\right)=-\frac1a\,\left(1+\frac{\dot H}{H^2}\right)=-\frac{1}{(aH)^2}\,\ddot a\quad.
\eeq
An early phase of \emph{decreasing} comoving horizon requires an interval of \emph{accelerated} expansion $\ddot a >0$ ending at $t=t_e$, and this is inflation in its most general sense. 

By inspecting the last result, we conclude that
\beq
\frac{\ddot a}{a}=H^2 \,\left(1+\frac{\dot H}{H^2}\right)= H^2\,(1-\epsilon_H)\quad{\rm with}\quad \epsilon_H\equiv-\,\frac{\dot H}{H^2}\quad.
\eeq
Acceleration implies a condition on the \emph{first slow-roll parameter} $\epsilon_H$, namely we need $\epsilon_H <1$. While this is the general condition for inflation, we see a very simple means of realizing this regime: We look for an energy source yielding $\dot H \approx 0$ in the sense of $-\dot H \ll H^2$ (since inflation must end at $t=t_e$, we cannot have $\dot H = 0$ strictly). This implies \emph{Hubble slow-roll} $H\approx const.$ and leads to a quasi-exponential phase of expansion $a\sim e^{Ht}$. Thus, building inflation in the more narrow sense means constructing a fluid which behaves almost like vacuum energy with $w\approx -1$ for a while and then quickly changes towards $w\geq 0$ at $t=t_e$. 

We now see that inflation driven by a source similar to vacuum energy with $\rho\approx const.$ solves the flatness problem as well. The contribution of spatial curvature $-k/a^2$ to the Friedmann equation shrinks exponentially during inflation compared to the source $\rho\approx const.>0$ driving inflation. Hence, if the exponential expansion latest long enough, inflation can render the universe spatially flat enough at the beginning $t_e$ of the matter or radiation driven expansion to avoid spatial curvature growing to more than about a percent fraction now.

Plugging in the ratio of scales between, in the most extreme case, the GUT scale and the largest cosmological scales visible today, we see that inflation must grow the scale factor by at least
\beq
a(t_e)\simeq a(t_\star)e^{60}
\eeq
60 efolds, to suppress spatial curvature to less than a percent in our present late-time universe. Solving the horizon problem, i.e. requiring that inflation lasted long enough to have all the scales between the GUT scale and cosmological scales today inside the horizon at the beginning $t_\star$ of the needed amount of inflation, needs again about 60 efolds of inflation.

A very large class of inflation models realizing this path uses the dynamics of one or several scalar fields. A scalar field minimally coupled to Einstein gravity is described in 4D at the 2-derivative level by an action
\beq\label{eq:scalaraction}
S=\int d^4x\sqrt{-g}\left[\frac12R+\frac12 (\partial_\mu \phi)^2-V(\phi)\right]\quad.
\eeq
If we can arrange for a regime where $V >0$ and $\dot\phi^2 \ll V$, then the scalar potential will act effectively as a positive vacuum energy and drive exponential expansion. The condition
\beq
\dot\phi^2 \ll V
\eeq
guarantees 
\beq
\epsilon_H\ll 1\quad,
\eeq
the first Hubble slow-roll condition for inflation. In this case, we can also further express $\epsilon_H\ll 1$ by a 1st slow-roll condition on the scalar potential itself
\beq
\epsilon_H\simeq \epsilon\equiv \frac12\,\left(\frac{V'}{V}\right)^2\ll1
\eeq
where we denote $()'=\frac{d}{d\phi}()$. Maintaining this condition for a long time to generate at least about 60 efolds of exponential expansion, then requires a 2nd slow-roll condition on the scalar potential to hold
\beq
\eta\equiv\frac{V''}{V}\ll 1\quad.
\eeq
Scalar field models of inflation realizing Hubble slow-roll this way are called slow-roll models of inflation~\cite{LINDE,STEINHARDT}.

Alternatively, higher-derivative kinetic terms can generate a phase $\epsilon_H \ll 1$ even if $V=0$ or if $\epsilon,\eta>1$ for the potential itself. One example of this in field theory is $k$-inflation~\cite{KINF}, and a simple realization of the same higher-derivative mechanism to generate Hubble slow-roll in string theory is DBI-inflation~\cite{DBI}.

\subsection{Quantum fluctuations during inflation}\label{sec_infquant}

The description of inflation using the dynamics of a scalar field has a very far-reaching consequence. We have seen, that successful slow-roll inflation entails the scalar inflaton being light $\eta \ll 1 \Leftrightarrow m_\phi^2 << H^2$. However, light scalar degrees of freedom are subject to quantum mechanical vacuum fluctuations. Inflation takes these fluctuations and stretches their wavelength so rapidly, that they become larger than the Hubble horizon ('super-horizon') after a finite time. The finite average amplitude, which a given fluctuation has at that point, then 'freezes' since super-horizon wave-length fluctuations cease to evolve during inflation. They have become an essentially classical field profile at this stage. Once inflation ends, these frozen large wave-length modes re-enter the horizon at a certain time given as a function of their comoving wavenumber. The modes who left the horizon during inflation first, re-enter last after inflation. The scalar field distribution described by these long wave-length modes causes a variation of the gravitational potential. This 'curvature perturbation' generates an initial field of density perturbations in the matter distribution after inflation. Hence, in this picture the inflationary quantum fluctuations are the ultimate cause of the primordial density perturbations which are the seed of structure formation in the observable universe~\cite{CHIBISOVMUKHANOV}.

We describe the curvature perturbation $\zeta$ by expanding around the FRW metric
\beq
ds^2=N^2 dt^2-h_{ij}\left(dx^i+N^idt)(dx^j+N^jdt\right)
\eeq
where $N, N^i$ denote the lapse function and shift vector, respectively, which enforce constraints containing the gauge invariance of GR. The spatial metric reads
\beq
h_{ij} = a^2(t) \left(e^{2\zeta}\delta_{ij}+\gamma_{ij}\right)\quad.
\eeq
During inflation we have $a(t)\approx e^{Ht}$.
Plugging this ansatz into the action eq.~\eqref{eq:scalaraction} and expanding in $\zeta$ and the inflaton fluctuation $\delta\phi$ leads at leading order to a 2-derivative action for $\zeta$ and $\delta\phi$ without self-interactions. Time reparametrization invariance relates $\zeta$ and $\delta\phi$. This gives us the freedom to trade between $\zeta$ and $\delta\phi$ and implies the relation
\beq
\zeta=\frac{H}{\dot\phi}\delta\phi\quad.
\eeq
We usually characterize the fluctuation spectrum of $\zeta$ in terms of the power spectrum $\Delta_\zeta^2$, which is the Fourier transform of the 2-point function $\langle\zeta(\bf x)\zeta(\bf y)\rangle$. We define the power spectrum via
\beq
\langle A_k A_{k'}\rangle\equiv \frac{\Delta_A^2}{k^3}\delta(\bf k+{\bf k'})
\eeq
for a fluctuation field $A(\bf x)$. The relation between $\zeta$ and $\delta\phi$ implies
\beq
\langle\zeta_k\zeta{k'}\rangle=\frac{H^2}{\dot\phi^2}\langle \delta\phi_k\delta\phi_{k'}\rangle\quad.
\eeq
In slow-roll inflation we have 
\beq
\langle \delta\phi_k\delta\phi_{k'}\rangle=\frac{1}
{k^3}\left(\frac{H}{2\pi}\right)^2\delta(\bf k+{\bf k'})\quad,
\eeq
and so we get
\beq
\langle \zeta_k\zeta_{k'}\rangle=\frac{H^4}{4\pi^2\,\dot\phi^2\,k^3}\,\left(\frac{k}{k_\star}\right)^{n_s-1}\delta(\bf k+{\bf k'})\quad.
\eeq
Here $k_\star$ defines a reference wavenumber corresponding to large-scale CMB fluctuations (e.g. corresponding to $\ell_\star=200$ after decomposition of the wave field into spherical harmonics). After the magnitude $\Delta_\zeta^2$ of the curvature perturbation power spectrum, the spectral tilt $n_s$ constitutes our 2nd inflationary observable. $n_s$ captures the 1st-order variations of the slowly rolling scalar field and is calculable for any given model of inflation.

Generically, interactions of the inflaton will generate higher-point functions $\langle \zeta_{k_1}\cdots\zeta_{k_n}\rangle$ from expanding out eq.~\eqref{eq:scalaraction} beyond 2nd order. Non-vanishing odd-point functions, such as the 3-point function $\langle \zeta_k\zeta_{k'}\zeta_{k''}\rangle$, constitute 'non-Gaussianity'. Two central results establish that non-Gaussianity is small for single-field slow-roll inflation, while the magnitude of non-Gaussianity is linked to the 'speed of sound' $c_s$ of the curvature perturbation. Furthermore, there are certain relations between the 'shapes' in momentum space which the 3-point function can take~\cite{MALDACENANG,EFTNG,Ade:2013ydc}. For instance, DBI inflation~\cite{DBI} produces a distinctive pattern of non-Gaussianity, where the 3-point function peaks in the 'equilateral' configuration, where all three momenta have roughly equal magnitude. This equilateral shape is not yet that strongly constrained from the PLANCK data $f_{NL}^{equil.}=42\pm75$, contrary to the local shape from multi-field inflation $f_{NL}^{loc.}=2.7\pm5.8$~\cite{Ade:2013ydc}.

Finally, there are gravitational waves in general relativity. These appear as the fluctuations $\gamma_{ij}$, hence called 'tensor modes'. Each polarization $s$ of a gravitational wave formally constitutes a massless scalar degree of freedom. Hence, inflation generates a long wave-length spectrum of gravitational waves with a power spectrum
\beq
\sum\limits_{s,s'=1,2}\langle \delta\gamma_{k,s}\delta\gamma_{k',s'}\rangle\equiv\frac{\Delta_T^2}{k^3}\delta({\bf k}+{\bf k'})=\sum\limits_{s,s'}\frac{1}
{k^3}\left(\frac{H}{\pi}\right)^2\delta{ss'}\;\delta(\bf k+{\bf k'})\quad.
\eeq
Observationally, we often refer instead to the 'tensor-to-scalar ratio' $r$ of gravitational wave power to curvature perturbation power
\beq
r\equiv \frac{\Delta_T^2}{\Delta_\zeta^2}=8\frac{\dot\phi^2}{H^2}\quad.
\eeq
From the PLANCK~\cite{Ade:2013zuv,Ade:2013uln,Ade:2013ydc} and WMAP~\cite{Hinshaw:2012fq} satellite data, we know $\Delta_\zeta^2\simeq 2.2\times 10^{-9}$. Hence, any detection of $r\gtrsim 0.01$ implies a GUT-scale inflaton potential $V_{inf.}^{1/4}\sim M_{GUT}\sim 10^{16}\,{\rm GeV}$ in the context of single-field slow-roll inflation. This is one of the main reasons driving the search for tensor modes.

Tensor modes are detectable in the CMB as a B-mode polarization pattern. Recently, the BICEP2~\cite{BICEP2} experiment reported on the detection of B-mode signal in the CMB on large angular scales $\ell\sim 100$. If future analyses along the line of~\cite{DUST1,DUST2} using e.g. PLANCK polarization data confirm a part of this B-mode signal to be primordial as opposed to coming from polarized galactic dust, then we would know that $r\sim 0.1$.

In models of single-field slow-roll inflation we can compute $n_s$ and $r$ as functions of the slow-roll parameters. To leading order we get
\beq
n_s=1-6\epsilon+2\eta\quad,\quad r = 16\epsilon\quad.
\eeq
We can express these results in terms of the number of e-folds $N_e(\phi)$ that inflation lasts from the field value $\phi$ until the end of inflation at $\phi_e$. We have
\beq
N_e=\int\limits_{t(\phi_e)}^{t(\phi(N_e))}H\,dt=\int\limits_{\phi_e}^{\phi(N_e)}\frac{d\phi}{\sqrt{2\epsilon}}\quad.
\eeq
CMB scales correspond to about $N_e=50\ldots 60$ e-folds before the end of inflation.
If we assume $\epsilon$ to increase monotonically with $\phi$ then we can bound $N_e$ by
\beq
N_e<\frac{\Delta\phi}{\sqrt{2\epsilon}}\quad.
\eeq
This implies the Lyth bound~\cite{LYTH}
\beq
\Delta\phi> \sqrt{\frac{r}{8}} N_e\quad.
\eeq
For $N_e\simeq 50\ldots 60$ we see that $r>0.01$ implies $\Delta\phi > M_{\rm P}$. This ties a large B-mode signal from primordial tensor modes to having a so-called 'large-field model' of inflation, where sufficient inflation requires a trans-Planckian initial field displacement.

\subsection{Effective field theory and the role of symmetries}

We now need to discuss inflation in effective field theory, the role which the amount of field displacement $\Delta\phi$ plays for the ultraviolet (UV) sensitivity of inflation. In effective field theory in the Wilsonian sense we Taylor expand the effective action of the inflaton scalar field around a given point $\phi_0$ in field space. We can organize this expansion in terms of the scaling of the various operators under dilatations of space and time $x^\mu\to ax^\mu$. The operator dimension of $\phi$ follows from the requirements of an invariant kinetic term, so $\phi\to\phi/a$. This is necessary for the action to be dimensionless as it is the phase in the path integral. Expanding the non-kinetic part of the action we get (assuming a $Z_2$ symmetry $\phi\to-\phi$ for slight simplification)
\beq
\begin{split}
S=\int d^4\sqrt{-g}& \left[ \frac12\left(\partial_\mu\phi\right)^2-\frac{m^2}{2}(\phi-\phi_0)^2+\sum\limits_{n\geq0}\frac{\lambda_{4+2n}}{M_\star^{2n}}(\phi-\phi_0)^{4+2n}\right.\\
&\left.\qquad+\frac{\lambda_{4,4}}{M_\star^4}\left(\partial_\mu\phi\right)^4+\ldots\right]\quad.
\end{split}
\eeq
Except for the mass all couplings are dimensionless as the relevant mass scale $M_\star$ suppressing the higher-dimension contributions appears everywhere. The higher-dimension operators scale under dilatation as ${\cal O}\to\frac{1}{a^\Delta}{\cal O}$ where e.g. for the non-derivative operators above we have $\Delta_{4+2n}=4+2n$ at weak coupling. We see that these operators induce scattering amplitudes with effective couplings $\lambda_{4+2n, eff.}=\lambda_{4+2n} (E/M_\star)^{\Delta_{4+2n}-4}$. Hence, operators with $n<0$, such as the mass term, are 'relevant' at low energies. The quartic self-coupling of a scalar ($n=0$) is an example of a 'marginal' operator. The great majority of operators have $n>0$. These are 'irrelevant' as they die out in the infrared. In the following discussion we will typically have the UV mass scale $M_\star=M_{\rm P}$

However, the viability of inflation depends on the slow-roll parameters being small $\epsilon,\eta\ll1$. As they contain $\phi$-derivatives, we see that in particular $\eta$ which pulls down two powers of $\phi$ can receive large contributions from 'dangerously irrelevant' operators ${\cal O}_{4+2n},n>0$. In particular, an operator of the form ${\cal O}_6=\lambda_6 V_0(\phi)(\phi-\phi_0)^2/M_{\rm P}^2$ corrects $\eta$ by an ${\cal O}(1)$ shift which destroys inflation. This happens  even at low energy densities during inflation $H/M_{\rm P}\ll1$ and for any field displacement $\Delta\phi=\phi-\phi_0$.

Moreover, we see that for $\Delta\phi\gg M_{\rm P}$ \emph{all} higher-dimension operators of type ${\cal O}_{4n+2}$ above correct $\eta$ by ${\cal O}(1)$ values, while for $n\geq2$ their contribution is suppressed in $(\Delta\phi/M_{\rm P})^{2n-2}\ll1$ at small $\Delta\phi\ll M_{\rm P}$. This is the true significance of the split into 'small-field' $\Delta\phi\ll 1$ and 'large-field' $\Delta\phi$ models of inflation.

We see that viable large-field inflation requires extra suppression of all higher-dimension corrections $\lambda_{4-2n}\ll1$. This amounts to the presence of a symmetry effectively forbidding these terms at the high scale. As the primary inflationary scalar potential itself has $V(\phi)\ll 1$ this symmetry effectively takes the form of a shift symmetry $\phi\to\phi+c$. An exact shift symmetry forbids non-derivative couplings. We can now argue that breaking the shift symmetry \emph{weakly and smoothly} by a large-field inflaton potential $V(\phi)\ll 1$ is sufficient to protect against the dangerous generic higher-dimension corrections discussed above. The arguably simplest large-field (and likely simplest overall) inflaton potential is a mass term~\cite{LINDECHAOS}
\beq
V(\phi)=\frac12m^2\phi^2\quad.
\eeq
The ensuing argument works the same for a direct generalization to monomial potentials $V(\phi)=\mu^{4-p}\phi^p$, $p>0$.

Firstly, we see that expanding the potential around a reference point $\phi_0$ we get
\beq
V(\phi)=V(\phi_0)\left(1+\frac{\delta\phi}{\phi_0}\right)^p=V(\phi_0)+\sum\limits_n\left(\begin{array}{c}p \\ n\end{array}\right)\frac{\delta\phi^n}{\phi_0^{n-p}}\quad.
\eeq
During inflation $\phi_0\gg1$ and so we see that the interaction terms $n\geq 3$ have effective self-couplings of the inflaton dying out as $1/\phi_0^{n-p}$ at large field displacements. We therefore expect the potential to be safe from dangerous radiative corrections induced by self-interactions at large field values.

Secondly, we can see this in field theory if we look at the relevant Feynman diagrams.\cite{KALOPER08,KALOPER11}. These daisy diagrams individually produce catastrophic-looking contribution, but their sum constitutes an \emph{alternating} series, as e.g. for $\phi^4$ theory
\begin{figure*}[h]
\centerline{\psfig{file=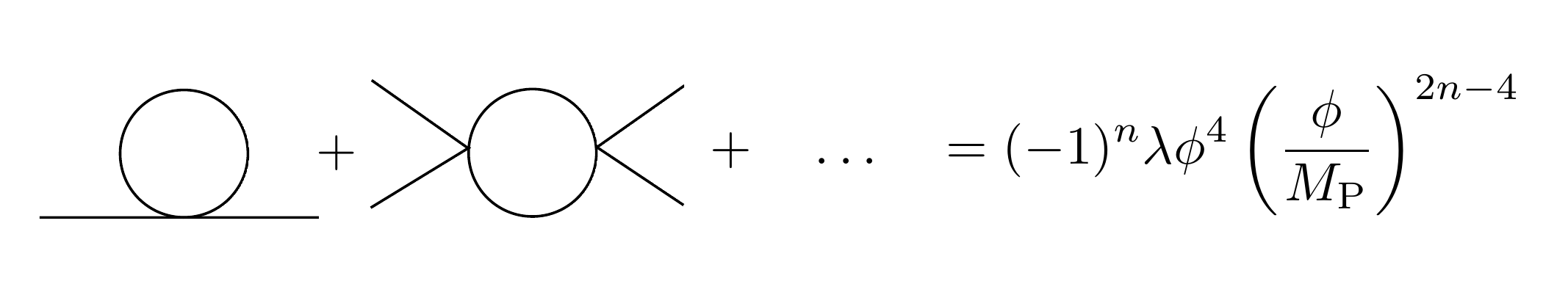,width=11cm}}
\end{figure*}
They resum into a good-natured logarithmic correction
\beq
V(\phi)=\lambda\phi^4\left[1+c\ln\left(\frac{\phi}{M_{\rm P}}\right)\right]
\eeq
which corrects the numerical values of $n_s, r$ a bit but is far from spoiling slow-roll.

Finally, quantized general relativity as a low-energy limit of quantum gravity couples \emph{only} to $T_{\mu\nu}$. This is sourced by $V(\phi)$ and $\partial_{\phi}^2V$, but \emph{not} by field displacements themselves. Consequently, graviton loops induce corrections to the effective scalar potential and Newton's constant of the form~\cite{SMOLIN,LINDEBOOK,KALOPER11}
\beq
\delta V\sim \left(c\frac{V}{M_{\rm P}^4}+d\frac{\partial_\phi^2V}{M_{\rm P}^2}\right) V(\phi)\quad,\quad \delta{\cal L}_{EH}\sim \sqrt{-g}\,\partial_\phi^2V R\quad.
\eeq
Large-field inflation requires $\mu\ll1$ and hence $V,\partial_\phi^2 V\ll 1$ to realize a COBE-normalized CMB spectrum from the curvature perturbation. Hence, all quantum gravity corrections are highly suppressed in $V/M_{\rm P}^4\;,\; \partial_\phi^2 V/M_{\rm P}^2 \ll 1$ except at extremely large field displacements where $V/M_{\rm P}^4\;,\; \partial_\phi^2 V/M_{\rm P}^2 \lesssim 1$.

The upshot of the whole discussion is that large-field slow-roll inflation is technically natural in t'Hooft's sense. If we succeed to generate the mass scale $\mu\ll M_{\rm P}$ of a given large-field model dynamically, then it would become even natural in the Wilsonian sense.

\subsection{Need for UV completion}

From the above discussion it is clear that large-field inflation works perfectly fine in effective field theory. However, the role of a UV complete candidate theory of quantum gravity such as string theory becomes clear: Firstly, the theory has to describe how the shift symmetry of the inflaton arises in the first place. Secondly, the UV completion needs to describe how it generates the large-field inflaton potential including the large trans-Planckian field range. Finally, a fundamental description needs to describe how the coupling of the inflaton to heavy states such as the moduli of string theory participates in the breaking of the shift symmetry and backreacts on the dynamics of slow-roll.

We will discuss in detail below, how shift symmetries of axions in string theory arise as remnants of its higher-dimensional gauge symmetries. Hence, these  $p$-form axions as natural large-field inflaton candidates in string theory. However, string theory seems to limit the periodicity field range of such axions to sub-Planckian values\footnote{These bounds are particularly strict for closed-string sector axions. For axions from the open string sector we may have slightly less restrictive situations, see e.g.~\cite{Kenton:2014gma}}. We will see that monodromy induced by the presence of higher $p$-form fluxes and/or branes breaks the periodicity generating a large-field range monomial potential for the axions. Combined with moduli stabilization this can be done in a regime $V(\phi)\ll1$ which provides radiative stability as seen above.

String theory provides the setting to discuss the interaction between the many moduli of its myriad compactifications to 4D and the inflationary dynamics of shift symmetry breaking from monodromy. However, we can see the basic effect of integrating out the heavy moduli already at this point by coupling a large-field inflaton to a heavy modulus field in field theory~\cite{DHSW}
\beq
V(\phi,\chi)=\frac12 \phi^2\chi^2+\frac12 M_\chi^2(\chi-\chi_0)^2\quad.
\eeq
If we choose $\chi_0\sim m\ll 1$ and a heavy modulus $M_\chi \gg m$ we see that this generates $V\sim m^2\phi^2$ as the effective potential at small $\phi$. However, at larger values of $\phi$ the full effective potential after integrating out $\chi$ reads
\beq
V_{eff.}(\phi)=M_\chi^2m^2\frac{\frac12\phi^2}{\phi^2+M_\chi^2}\quad.
\eeq
Hence the presence of the heavy field leads to a \emph{flattening} of the inflaton potential $V(\phi)\sim \phi^p$ compared to its naive tree-level power $V_0(\phi)\sim \phi^{p_0}$. This behavior is generic and occurs in many cases of large-field axion monodromy inflation after carefully including moduli stabilization~\cite{DHSW,KHW,MSWW}.

\section{Ingredients of String Theory}\label{sec_strings}

The above discussion gives us a clear motivation to use string theory as a UV completion of inflation. We do not yet know if string theory is the correct description of our world. However, it is a highly successful candidate theory of quantum gravity while including a unification of SM-like gauge forces, chiral fermion matter. Strong supporting evidence for this claim arises from the various mathematical consistency checks and 'null' results obtained in the last 20 years or so. These include the web of dualities linking all five string theories and 11D supergravity, the AdS/CFT correspondence which provides a setting where a certain non-perturbative definition of quantum gravity is available, and the description of black hole entropy (even though for highly supersymmetric settings only so far).

We do not have a complete picture or non-perturbative description of string theory yet. Given that we have so far only various corners where perturbative control and/or the use of string dualities allow us calculational access, we cannot yet compare the theory as a whole with experimental data. At this stage we can only try to identify mechanisms in string theory which realizes certain types of dynamics which we need to describe our world. Certain mechanisms which realize inflation in controlled string theory settings will be our prime examples here. These mechanisms may then serve for comparisons with data, or they may inspire new model building and analysis for the bottom-up construction of inflation models in effective field theory as well.

\subsection*{The moving parts}

Given this preamble, we start with collecting the necessary ingredients of string theory which guide our current understanding of the many vacua we get from string compactification to 4D. These ingredients provide the necessary dynamics for moduli stabilization, and building string inflation within the stabilized vacua of the string landscape. The 2D worldsheet action of superstrings gives rise to a 10D target spacetime if we demand the Weyl anomaly to vanish for a flat target spacetime. Note, that giving up flatness of the target space allows for string theories with $D>10$ which have interesting consequences such as a spectrum with $2^D$ axions~\cite{EVAnoncrit,2toDaxions1,2toDaxions2,2toDaxions3}. The maybe simplest exact solution then is 10D Minkowski space with either ${\cal N}=1$ or $2$ supersymmetry. Hence, in 10D the effective spacetime action is one of the five possible 10D supergravities (type I, heterotic $SO(32)/E_8\times E_8$, type IIB, type IIA). We can write these 10D actions schematically as~\cite{Polchinski:1998rr}
\beq
S=\frac{1}{2\alpha'^4}\int d^{10}x\sqrt{-G}\,e^{-2\phi}\left(R+4(\partial_M\phi)^2\right)+S_{matter}
\eeq
The matter part needs more detailed discussion. We can write its action, again schematically, as
\beq\label{eq:Smatter}
\begin{split}
S_{matter}=\int d^{10}x\sqrt{-G}&\left[e^{-2\phi}|H_3|^2+\sum\limits_p |\tilde F_p|^2+C.S.\right.\\
&\left.\quad+\sum\limits_p\left( T_p^B\frac{\delta^{9-p}(x_\perp)}{\sqrt{-G_\perp}}-T_p^O\frac{\delta^{9-p}(x_\perp)}{\sqrt{-G_\perp}}\right)\right.\\
&\left.\quad+higher-deriv.\right]
\end{split}
\eeq
Here C.S. denotes Chern-Simons terms involving the various $p$-form gauge potentials and field strengths of string theory. The terms localized at the $9-p$ coordinates $x_\perp$ denote $p$-branes with tension $T_p^B$ and orientifold planes with tension $T_p^O$. These different terms contribute various forms of potential energy to the effective action. The various branes labeled by $T_p^B$ contribute effectively terms proportional to their tension times their world-volume they wrap in the internal dimensions as potential energy. On a subclass of them, called D-branes, we can have open strings ending on them. This renders D-branes dynamical. Their positions in the orthogonal dimensions are scalar fields leaving on the world-volume of a D-brane. Branes can carry charge under some of the higher-dimensional $p$-form gauge potentials of string theory, and we will make extensive use of that.

$T_p^O$ labels objects with negative tension. These are so-called orientifold planes (O-planes) which arise from modding both the internal space and the spectrum with generalizations of $Z_2$ action. In general, the number and type of allowed O-planes follows from certain topological properties and quantum numbers of the compact six dimensions. Hence, the theory limits the amount of negative energy objects which is rather beneficial for vacuum stability.

The various terms $|H_3|^2$ and $|\tilde F_p|^2$ denote the kinetic and gradient energy contributions from the field strengths of higher-dimensional $p$-form gauge potentials in string theory. They also contribute what is known as quantized background fluxes to the effective action. Since we will employ these objects far and wide, we will start by discussing their properties and utilize their relation with ordinary electrodynamics as a theory of a 1-form gauge potential.

Our discussion will follow the lines of e.g.~\cite{EVA2013TASI}. The electromagnetic gauge potential $A_1=A_{1,\mu}dx^\mu$ is a useful means of describing electromagnetism though containing redundancy. The field strength $F_2=dA_1$, $F_{2,\mu\nu}=\partial_\mu A_{1,\nu}-\partial_\nu A_{1,\mu}$ is gauge invariant under the gauge transformation $A_1\to A_1+d\Lambda$. This is not the only gauge invariant object object, as there are for instance Wilson lines $e^{i\int A_1}=e^{i\oint A_{1,\mu}dx^\mu}$ around a compact extra dimension $y$. Let us look at the 5D metric with the $y$-direction compactified on a circle $S^1$ with radius $L_y/(2\pi)$ (so $y\to y+1$)
\beq
ds^2=dt^2-dx_3^2-L_y^2 dy^2\quad.
\eeq
There is a flat gauge connection $A_1=A_{1,y}dy$ constituting a Wilson line around the $S^1$ such that
\beq
A_1=a(x)\omega_1\quad,\quad \omega_1=dy\quad,\quad a(x)\equiv \oint A_{1,y}dy\quad.
\eeq
The quantity $a(x)$ with periodicity $a\to a+1$ inherited from the $S^1$ is effectively an axion in 4D with a perturbatively exact shift symmetry. $a(x)$ is gauge invariant under $A_{1,y}\to A_{1,y}+\partial_y\Lambda$ since an $S^1$ has no boundary. Reducing out the gauge kinetic term for $F_2$ we get
\beq
\begin{split}
S_{F_2}=&\int d^4xdy \sqrt{-g}|F_{2,mn}|^2= \int d^4x\sqrt{-g_4}\underbrace{\int dy}_{=1}\underbrace{\sqrt{-g_{yy}}}_{L_y} g^{00}\underbrace{g^{yy}}_{1/L_y^2} |F_{2,0y}|^2\\
=&\int d^4x\sqrt{-g_4}L_y \frac{\dot a^2}{L_y^2}
\end{split}
\eeq
The axion kinetic term appears with an overall prefactor $f^2$, where we call $f$ the axion decay constant which here is $f=1/\sqrt{L_y}$, and thus inversely proportional to a power of the compact length scale. This is a generic property of all of the known string theory axions which arise by close analogy from the higher $p$-form gauge potentials in string theory in the presence of compact extra dimensions.

We now couple the electromagnetic field to a point charge. The interaction term
\beq
\int d^4xJ^\mu A_{1,\mu}=\int\limits_{worldline}A_1
\eeq
effectively describes the Aharonov-Bohm effect: the world-line integral picks up a phase along a closed path around a region containing a $B$-field, i.e. magnetic flux $\int {\bf B}d{\bf A}$. This explains a second property of magnetic fluxes: on a compact space they are quantized, for instance $\int_{S^2} F_2=N$. This should better be, as otherwise the Aharonov-Bohm phase picked up by a particle around a small loop on the compact space would not be a multiple of $2\pi$ producing a multi-valued wave function.

All of these properties generalize to higher dimensions with higher $p$-form gauge potentials, where the objects charged under the gauge fields generalize then as well from point charges to charged branes. The first basic generalization comes from the anti-symmetric 2-form gauge potential $B_2$ for which the Polyakov action of the string allows a coupling on the string world-volume $\Sigma$ reading
\beq
S_{string}\supset \int_\Sigma B_2=\frac{1}{\alpha'}\int d^2\sigma \sqrt{-\gamma} \epsilon^{ab}B_{2,MN}\partial_a X^M\partial_b X^N\quad.
\eeq
The 10D effective action for $B_2$ is the $|H_3|^2=|dB_2|^2$ term in eq.~\eqref{eq:Smatter} above, with the field strength of $B_2$ being $H_3=dB_2$. The form of $H_3$ provides for gauge invariance under $B_2\to B_2+d\Lambda_1$. In close analogy with the electromagnetic case before this leads to axions
\beq
b_i(x)=\int_{\Sigma_2^i}B_2\quad,\quad B_2=b_i(x)\omega_2^i
\eeq
on non-trivial trivial 2-cycles $\Sigma_2^i$ with their associated basis 2-forms $\omega_2^i$. Again, the wavefunction of a string picks up an Aharonov-Bohm like phase on a loop around a region with magnetic flux $H_3$ due to the $B_2$ world-volume coupling above. Consequently, there are quantized background fluxes $\int_{\Sigma_3^a}H_3=N_a$ of $H_3$ on non-trivial 3-cycles $\Sigma_3^a$ of the compact dimensions.

The first natural guess to get inflation from the $b$-axion would be to implement natural inflation~\cite{NATINF}. The kinetic term for $b$-axion arises from reducing $|H_3|^2$, while a Euclidean string worldsheet wrapping an appropriate 2-cycle $\Sigma_2$ of the extra dimensions may provide a non-perturbative contribution to the scalar potential of $b$. We get
\beq
{\cal L}=f^2(\partial_\mu b)^2-\Lambda^4\,\left[1-\cos(b)\right]\quad.
\eeq
Canonically normalizing $b$ into $\phi=bf$ gives the potential
\beq
V=\Lambda^4\,\left[1-\cos(\phi/f)\right]\quad.
\eeq
For large $f\gg M_{\rm P}$ the potential approximates $m^2\phi^2$ around the minimum.
We can now compute $n_s$ and $r$ from the slow-roll parameters. Then, we find that requiring $n_s > 0.945$ as required by the PLANCK 95\% confidence limits implies $f\gtrsim 4.5 M_{\rm P}$~\cite{AXIONINFREV,STRINGINF}. However, the $b$-axion kinetic term gets a prefactor $f^2\sim M_{\rm P}^2/L^q$ with $q>1$ for a 2-cycle with length scale $L$ in string units. This is in full analogy with the simple electromagnetic case before. Since a controlled string compactification requires all radii to be large, we have $L\gg 1$ and hence $f< M_{\rm P}$. This behavior is generic -- all known string axions both from NSNS-sector or RR-sector $p$-form gauge fields have sub-Planckian axion decay constants~\cite{Gorbatov}. Attempts to evade this were so far forced to resort to small radii or large string coupling~\cite{BLUPLAU14,GRIMM14}.

\section{Axion Monodromy}\label{sec_stringinf}

Pushing further we note that in analogy with $B_2$ there are RR-sector $p$-form gauge potentials $C_{p-1}$ and their field strengths $F_p=dC_{p-1}$. They arise from the type I/type II open string sector and have various branes as objects whose world-volume couples to $C_{p-1}$. By the same arguments as above, RR-sector axions $c^{(p-1)}_k(x)=\int_{\Sigma_{p-1}^k}C_{p-1}$ arise from the invariance of $F_p$ under $C_{p-1}\to C_{p-1}+d\Lambda_{p-2}$.

Finally, the full duality structure and set of gauge invariances of the various string theories forces generalization of the RR-sector $p$-form field strengths to include various Chern-Simons couplings
\beq
\tilde F_p=F_p+B_2 \wedge F_{p-2}\quad.
\eeq
The corresponding kinetic terms $|\tilde F_p|^2$ are invariant under $B_2$ gauge transformation as well, if at the same time $C_{p-1}\to C_{p-1}-\Lambda_1\wedge F_{p-2}$.

\subsection{Axion monodromy inflation}

We see that turning on RR-flux $F_{p-2}$ provides $C_{p-1}$ with a 'Stueckelberg' charge under the $B_2$ gauge transformation. $C_{p-1}$ now shifts with $\Lambda_1$, while at the same time $F_{p-2}$-flux provides a mass term from $|\tilde F_p|^2$ for the $B_2$ gauge field and its associated axion $b(x)$.

This is in close analogy to the phenomenon of superconductivity: there, spontaneous symmetry breaking leads to the appearance of a 'Stueckelberg scalar' $\theta(x)$ with a coupling $(A_1+d\theta)^2$ which provides the mass for the electromagnetic field inside a superconductor. The $A_1\wedge d\theta$ coupling is the analogue of the $B_2\wedge F_{p-2}$ coupling in higher dimensions. The coupling is necessary to preserve gauge invariance of the whole system even if the gauge field is massive, as now $A_1\to A_1+d\Lambda$ pairs with a shift $\theta\to\theta-\Lambda$.

Hence, turning on the lower-dimensional fluxes $F_{p-2}$ provides a \emph{non-periodic} potential for the $b$-axions of the form
\beq\label{eq:monodromicpot}
\begin{split}
V\sim &\int d^6y\sqrt{-g_6}|\tilde F_p|^2\\
\sim &\int d^6y\sqrt{-g_6}(F_p+B_2\wedge F_{p-2})^2\sim (N_p+b\; N_{p-2})^2\quad.
\end{split}
\eeq
Here $N_p$ and $N_{p-2}$ denote the flux quanta of $F_p$ and $F_{p-2}$ flux turned on. The field range of a given $b$-axion no longer shows periodicity and is a priori (kinematically) unbounded. This is the phenomenon of \emph{axion monodromy} parametrically extending the axion field range along a non-periodic potential from fluxes~\cite{MSW,KALOPER08,OSCILL,DHSW}. 

We recall here that the structure of the potential eq.~\eqref{eq:monodromicpot} is quite similar to the 4D field theory version of axion monodromy in~\cite{KALOPER08,KALOPER11,KALOPER14}. There, we start from an axion and a 4-form field strength
\beq
{\cal L}\sim (\partial_\mu\phi)^2+(F_{\mu\nu\rho\sigma})^2+\frac{\mu}{\sqrt{-g}}\phi\epsilon^{\mu\nu\rho\sigma}F_{\mu\nu\rho\sigma}\quad.
\eeq
Integrating out the 4-form while giving it $q$ units of flux gives $\phi$ a potential $V\sim (q+\mu\phi)^2$. The underlying shift symmetry appears as a joint shift of both $\phi$ and $q$, but shifts in $q$ are again mediated by exponentially suppressed brane nucleations. Hence picking $q$ chooses a branch, giving the axion $\phi$ a non-periodic, a priori quadratic potential. The structure of this 4D theory is rather similar to the reduction of $|\tilde F_p|^2$ above.

Note, that we can compensate integer shifts of the axion $b$ by appropriate integer changes in the $F_{p-2}$ flux quanta. However, these changes are mediated by non-perturbative effects which are suppressed at weak string coupling and large volume. Hence, the full system displays a set of non-periodic potential \emph{branches} for the $b$-axion~\cite{MSW,KALOPER08,KALOPER11,KALOPER14}. The branches are labeled by the flux quanta of $F_{p-2}$. The periodicity of the full theory is now visible in the set of branches of non-periodic potentials when summing over branches. However, as in spontaneous symmetry breaking, once we pick a certain quantized flux $F_{p_2}$ which can only change by exponentially suppressed effects, we pick a certain branch along which the $b$-axion rolls in a non-periodic potential. Hence, axion monodromy clearly lends itself to realize large-field inflation in the context of string theory.

The generic structure of the axion effective action on one such branch picked by $F_{p-2}$ flux looks like
\beq
{\cal L}=f^2(\chi)(\partial_\mu b)^2+\mu(\chi)^{4-p_0} b^{p_0}+\Lambda^4(\chi)\cos(b)\quad.
\eeq
$\chi$ summarily denote the moduli of a given string compactification, and the axion acquires a periodic contribution from non-perturbative contributions. However, we expect their scale $\Lambda^4$ to be exponentially suppressed in the radii of the extra dimensions and $1/g_s$. Hence, the generic axion potential is a large-field monomial with a priori power $p_0$ with tiny periodic modulations on top. The axion decay constant, however, is moduli dependent. As discussed above, in general $f\sim M_{\rm P}/L^q, q>0$. Hence, after moduli stabilization the backreaction of the moduli $\chi=\chi(b)$ due to the inflationary vacuum energy may very well change $f=f(\chi(b))$ and thus the canonically normalized inflaton field $\phi(b)$ we get from $d\phi(b)=f(\chi(b))db$.

Provided we realize this mechanism in a controlled setting with moduli stabilization at large volume and weak string coupling and with a sub-Planckian inflationary energy scale from axion monodromy, this mechanism will inherit all the properties of radiative stability discussed in the previous section. However, the presence of the moduli generically leads to backreaction effects as already discussed above in field theory. The CS-couplings in the generalized flux kinetic terms lead to a priori axion potentials $V\sim b^{p_0}$ with powers $p_0=2,3,4$~\cite{WEIG14,MSU,KHW,BLUPLAU14,GRIMM14,UTHROAT1,MSWW}. However, inclusion of the moduli will generically lead to \emph{flattening} from backreaction of the moduli onto the axion potential~\cite{DHSW,KALOPER14,MSWW}
\beq
V_{eff.}\sim \phi(b)^p\quad,\quad p<p_0\quad.
\eeq
The associated predictions for the inflationary observables are $n_s=1-(2+p)/(2N_e)$ and $r=4p/N_e\simeq0.05\ldots0.24$ for $p\simeq 0.5\ldots4$ and $N_e=50\ldots 60$.

Finally, we can see the same basic effect of axion monodromy arising from the coupling of the $p$-form gauge potentials to branes. This must be the case, as the web of dualities in string theory in the ends relates compactification with certain sets of branes to dual geometries without branes but background fluxes describing the same physics.

For simplicity, we look at the effective action of $Dp$-branes wrapping $p+1$ dimensions of space-time. To start, we look at the world-line action of a particle $S=\int dt\sqrt{1-\dot x^2}$. If this particle is charged under the electromagnetic gauge potential, then 
\beq
S=\int dt\sqrt{1-\dot x^2}+q\int\limits_{worldline}A_1
\eeq
This structure generalizes to higher dimensions, where $Dp$-branes by analogy have a world-volume action of DBI form together with a CS term providing charge under the various RR-sector gauge potentials
\beq
S_{Dp}=T_{Dp}^B\int d^{p+1}\xi\sqrt{-\det(G+B_2)-\alpha'F_2}+\mu_{Dp}^B \int \left(C_{p+1}+\ldots\right)\quad.
\eeq
$\det G$ here denotes the determinant of the induced metric on the world-volume of the brane, while $F_2$ is the field strength of a $U(1)$ gauge field which a single brane can contain.

Now let us take a 5-brane living on ${\cal M}_4\times T^2$. That is, the 5-brane fills all macroscopic 4D space-time, and wraps a small compact two-torus of the extra dimensions. Now we put a $B_2$ gauge field $B_{12}=-B_{21}=b(x)$ on the $T^2$, which in 4D is our $b$-axion. At same time there may be $N_2$ units of combined 2-form flux $N_2=\int_{T^2}{\cal F}_2$, where ${\cal F}_2\equiv B_2+\alpha' F_2$. Reducing the brane action plus the $|H_3|^2$ (for the $b$ kinetic term) to 4D, we arrive at
\beq
{\cal L}=f^2\dot b^2-\sqrt{vol(T^2)^2+(b+N_2)^2}\quad.
\eeq
Again, we see the appearance of an infinite set of potential energy branches labeled by the flux integer $N_2$. However, once we pick a flux $N_2$, we are on one given branch. The $b$-axion now acquires a non-periodic scalar potential which is linear
\beq
V_{D5}\sim b
\eeq
at large $b$ on each branch~\cite{MSW}. Doing this on a D7-brane instead can produce $V_{D7}\sim b^2$ instead~\cite{WEIG14}. On each branch $b$ will drive large-field slow-roll inflation, while relaxation to different branches with different $N_2$ is subject to generically exponentially suppressed brane nucleation tunneling events~\cite{KALOPER08,KALOPER11,KALOPER14}. The general story is the same as before, as in fact it is related to the flux-induced version above by duality.

\subsection{Inflating with several axions}

Since axions seem to be ubiquitous in string compactifications (they comprise roughly $h^{1,1}/2$ of the CY moduli, and for supercritical strings there are ${\cal O}(2^D)$ axions present in 4D~\cite{EVAnoncrit,2toDaxions1,2toDaxions2,2toDaxions3}), we might look for assistance effects of several axions helping each other towards slow-roll. The original example of this is N-flation\cite{Nflation,GRIMM2,NflationCic}, where $N\ll1$ axions are excited simultaneously from the minima of the non-perturbatively generated $V_i(\phi_i)=\Lambda^4[1-\cos(\phi_i/f)]$ cosine-potentials. The Hubble parameter adds up the squares of the displacements in the quadratic approximation to the axion potentials $H\sim \sqrt{\sum_i V_i}\sim \sqrt{\sum_i m^2\phi_i^2}\sim \sqrt{N}\sqrt{m^2\phi^2}\;,\;\phi_i\sim \phi \;\forall i$. This increases the friction term in the equation of motion for \emph{each} axion by a factor $\sim \sqrt{N}$. However, the Planck mass renormalizes as well, yielding in total
\beq
H^2\sim \frac{N\, m^2\phi^2}{M_{\rm P}^2+N\,M_\star^2}
\eeq
where $M_\star$ denotes the appropriate cut-off scale for the diagrams which renormalize the Planck scale. This renders parametric enhancement of $H$ at large $N$ difficult to achieve. However, we note here that describing the axion decay constants of an N-flation setup as a randomized ensemble enhances the relative likelihood of alignments or hierarchy among the decay constants. This can reduce the pressure towards large $N$ for N-flation somewhat~\cite{EastherMcAll,McAllNflat14}.

The other option analyzed tries to realize natural inflation using a 2-axion system. We start with a 2-axion system which is aligned in the axion decay constants such that it has precisely one flat direction~\cite{KNP,Kappl:2014lra}. We begin with
\beq
\begin{split}
{\cal L}=&f_r^2(\partial_\mu a_r)^2+f_\theta^2(\partial_\mu a_\theta)^2\\
&-\Lambda_1^4\,\left[1-\cos(p_1a_r+p_2a_\theta)\right]-\Lambda_2^4\,\left[1-\cos(q_1a_r+q_2a_\theta)\right]
\end{split}
\eeq
where the $p_i,q_i$ denote coefficients in the Euclidean action (e.g. coxeter numbers for gaugino condensation) of the nonperturbative effects generating the axion potential. Canonically normalizing the axions gives
\beq
V(r,\theta)=\Lambda^4_1\,\left[1-\cos\left(p_1\frac{r}{f_r}+p_2\frac{\theta}{f_\theta}\right)\right]+\Lambda^4_2\,\left[1-\cos\left(q_1\frac{r}{f_r}+q_2\frac{\theta}{f_\theta}\right)\right]\quad.
\eeq
Tuning an alignment $p_2/p_1 = q_2/q_1\equiv \kappa$ produces a single flat direction~\cite{KNP,Kappl:2014lra}. We have can now generate a shallow long-range axion potential by slightly perturbing the alignment. There are two ways to do so. One way is to manifestly tune a finite small misalignment into the above condition, changing it to $p_2/p_1=\kappa (1+\delta)$~\cite{KNP,Kappl:2014lra}. The flat direction lifts slightly producing an effective potential for the former flat direction 
\beq
V_{eff.}\sim 1-\cos(\phi_{eff.}/f_{eff.})\quad,\quad f_{eff.}=\frac{\sqrt{f_\theta^2+f_r^2\kappa^2}}{q_1\kappa}\,\frac{1}{\delta}\quad.
\eeq
For $\delta \lesssim 0.1$ the approximately flat direction has a large-field potential with $f_{eff.}> 5M_{\rm P}$ even if $f_r,f_\theta\lesssim M_{\rm P}$~\cite{KNP,Kappl:2014lra}. The smaller the initial decay constants $f_r$ and $f_\theta$ are, the smaller we must tune $\delta$ to reach a desired $f_{eff.}$. For 2 axions the near-alignment typically constitutes at least 1-10\% tuning. However, among a larger number of axions with a random decay constant distribution the likelihood for a random near-alignment of 2 axions with small enough $\delta$ among $N$ axions can be rather sizable~\cite{TAKAHASHI1}.

Alternatively, we can leave the aligned situation $V_0\sim 1-\cos(r/f_r+\theta/f_\theta)$ by simply adding a potential such that $V=W(r)+V_0(r,\theta)$. $W(r)$ can originate from some brane or flux induced monodromy~\cite{DI09}, or can come from a non-perturbative effect itself $W(r)\sim \cos(r/f'_r)$~\cite{BPW1,TYE14} (see also~\cite{Gao:2014uha}). Providing a hierarchy $f_r\ll f'_r,f_\theta$ is enough to give $\theta$ a long-range shallow potential with $f_{eff.}> 5M_{\rm P}$ for sub-Planckian initial decay constants. The potential for the inflaton, which is approximately $\theta$, sits like a set of terraces inside the potential slope for $r$.

\subsection{General structure of moduli stabilization}

String theory in its critical version lives in 10D. Describing four dimensional physics then typically employs compactification of the six extra dimensions on a small compact manifold. This process leaves various deformation modes as vacuum degeneracies which therefore describe massless and flat scalar fields in 4D, called the moduli. The presence of massless scalars is in contradiction with various experimental data (no 5th forces, light moduli screw up early Universe cosmology, etc.). The inclusion of background fluxes, branes, O-planes and internal curvature lead to a much improved understanding of moduli stabilization, that is giving mass to the various moduli~\cite{GKP}. Moduli stabilization also usually drives supersymmetry breaking. This enabled the first classes of de Sitter vacua in string theory, starting with the KKLT scenario~\cite{KKLT}.

Inflation in string theory is the phenomenon which after embedding in string theory potentially lives closest to the string scale of all observable phenomena so far (a verification of BICEP2~\cite{BICEP2} would imply $V_{inf.}^{1/4}\sim M_{GUT}\sim 10^{16}\,{\rm GeV}$, as discussed before). Given its dependence on a slow-rolling scalar field, moduli stabilization is absolutely unavoidable before a meaningful discussion of string inflation becomes possible.

Hence we sketch here the basics of moduli stabilization as they are relevant for our discussion. The various fluxes, branes, and O-planes, as well as internal curvature contribute after reducing to 4D potential energy terms which scale as inverse powers of the radii and the volume ${\cal V}$ of the extra dimensions, and as different positive powers of the string coupling $g_s$. These contributions fall into roughly three classes: i) certain branes, $H_3$-flux and negative internal curvature contribute positive terms of the form $V_1\sim +a g_s^{r_1}/{\cal V}^{q_1}$. ii) Negative terms $V_2\sim -b g_s^{r_2}/{\cal V}^{q_2}$ which fall slower ($q_2<q_1$) with inverse volume arise from O-planes, positive internal curvature, and some quantum corrections. iii) Finally, a 2nd set of positive terms falling even slower with the inverse volume, $V_3\sim +c g_s^{r_3}/{\cal V}^{q_3}$ arises from RR-sector fluxes. In total, these contributions add up to a scalar potential with a 3-term structure~\cite{EVAdS1,EVAdS2}
\beq
V=V_1+V_2+V_3= a \frac{g_s^{r_1}}{{\cal V}^{q_1}}-b \frac{g_s^{r_2}}{{\cal V}^{q_2}}+c \frac{g_s^{r_3}}{{\cal V}^{q_3}}\quad.
\eeq
Given the hierarchy $0<q_3<q_2<q_1$ so that the negative middle term can produce a dip in an other positive potential, tuning the fluxes and the negative term from O-planes and/or positive curvature allows to realize exponentially many dS vacua with suppressed cosmological constant. Due to the large number of fluxes, the number of these flux vacua, called the 'landscape', can easily surpass scales like $10^{500}$. This allows for environmental explanations of the observed amount of dark energy~\cite{KKLT,LENNY03,DENEFDOUGLAS,MODSTABreview}.

In general, stabilizing all moduli along the lines sketched above implies supersymmetry breaking at the KK scales, as turning on the required fluxes generically breaks SUSY~\cite{MODSTABreview}.

However, we may insist on preservation of a single 4D supersymmetry to low energies for various phenomenological reasons (such as the gauge hierarchy problem). This case pretty much forces us to compactify critical 10D string theory on a Calabi-Yau (CY) manifold, or an orientifold thereof in type II settings. Preserving the Calabi-Yau structure restricts the types of background fluxes which are admissible. In particular, we may only use imaginary self-dual 3-form flux $G_3=F_3-\tau H_3$ in type IIB compactifications which probably provide still the best understood examples of moduli stabilization so far~\cite{GKP}. $\tau=C_0+i/g_s=C_0+i e^{-\phi}$ denotes the axio-dilaton, the complexified type IIB string coupling.

The moduli space of a CY compactification decomposes into a set of $h^{2,1}$ 3-cycle complex structure moduli $U^a$ and $h^{1,1}_+$ K\"ahler or volume moduli $T_i$ which measure 4-cycle volumina in the CY. The index "+" refers to those volume moduli which are even under the orientifold projection needed for 4D ${\cal N}=1$ supersymmetry. A CY itself has $h^{1,1}\geq h^{1,1}_+$ volume moduli to start with. If $h^{1,1}> h^{1,1}_+$, the remaining $h_-^{1,1}$ moduli are orientifold-odd combinations of the $B_2$- and RR-sector $C_2$-form axions~\cite{GRIMM1,GRIMM2}, which in these settings are some of our natural large-field axion monodromy inflation candidates~\cite{MSW}. Turning on just the CY compatible 3-form fluxes results in 4D ${\cal N}=1$ effective supergravity for the moduli sector described by a K\"ahler potential and a superpotential
\begin{eqnarray}
K&=& -2\ln\,{\cal V}(T_i,
\overline{T_i})-\ln(-(\tau-\bar\tau)/2i)-K_{c.s.}(U^a,\overline{U^a})\nonumber\\
&&\\
W&=& W_0(U^a,\tau)\;\;,\quad W_0(U^a,\tau)=\int_{CY}G_3\wedge \Omega\quad.\nonumber
\end{eqnarray}
The 3-form fluxes induce a superpotential and corresponding scalar potential which serve to fix all the $U^a$ and axio-dilaton $\tau$~\cite{GKP,MODSTABreview}. However, this leaves the volume moduli $T_i$ with a no-scale scalar potential $V(T_i)=0\forall T_i$. 

It is this feature, which requires using a combination of perturbative $\alpha'$ and string loop corrections to the volume moduli K\"ahler potential~\cite{BBHL,LVS,LVSgs,LVSgs2}.
\beq
K_{\textrm{K\"ahler}}= -2\ln ({\cal V})\to K_{\textrm{K\"ahler}}= -2\ln ({\cal V}+\alpha'^3\xi)+\delta K_{g_s}
\eeq
and non-perturbative corrections in the $T_i$ from Euclidean brane instantons or gaugino condensation on D-brane stacks
\beq
W=W_0\to W=W_0+\sum_i A_i e^{-a_iT_i}
\eeq
to stabilize the volume moduli~\cite{KKLT,LVS,KAHLERUPLIFT,MODSTABreview}. These vacua are mostly AdS (except for those of \cite{KAHLERUPLIFT}). Reaching dS vacua involves either introducing manifestly SUSY breaking objects like anti-branes in warped regions (for control)~\cite{KKLT}, or effects generating F- and/or D-terms from the open string/matter sector~\cite{BKQ,LVSdS}. In these CY based constructions, the scale of SUSY breaking associated with volume stabilization tends to be lower due to the underlying no-scale structure than in more generic settings away from CYs. Some CY schemes such as the Large Volume Scenario (LVS)~\cite{LVS} can reach TeV scale gravitino mass either by stabilizing at very large volume or by using sequestering to suppress the soft masses with respect to an intermediate scale gravitino mass~\cite{LVSsequest}. We note, that in certain settings a racetrack combination of non-perturbative effects alone may suffice to stabilize all geometric moduli without the use of any flux. $G_2$ compactifications of M-theory provide one such an example~\cite{G2RT}.

The underlying no-scale structure thus forces us to stabilize the volume moduli at least partly by combinations of exponentially small effects. This feature of CY compactification tends to render the volume moduli more susceptible to backreaction from the inflationary vacuum energy. In turn, once we require CY compactification, this more often requires  us to separate the mass scale of the volume moduli from the scale of inflation, leading to more tuning in the moduli stabilization than necessary in the generic situation away from CYs~\cite{MSW,DHSW,KHW}. In contrast, in more generic settings with KK scale supersymmetry breaking, and all sources of moduli potential energy available in 10D in play, the inflationary vacuum often \emph{helps} with moduli stabilization~\cite{MSWW}. This leads to mostly harmless backreaction driving flattening of the inflaton potential, replicating the spirit of the field theory discussion of the last section.

We finally note, that~\cite{EVAdS1,EVAdS2,EVAdS3} used more general combinations of negative internal curvature, fluxes, branes and O-planes to generate new and more general classes of dS vacua by generalizing the basic flux and brane setup leading to the $AdS_5\times X^5$ setup of the AdS/CFT pairs. These constructions lead to very promising paths towards a holographic description of dS vacua in terms of $dS_d/dS_{d-1}$ pairs~\cite{EVAdSdS} or similar and potentially related FRW/CFT dual pairs~\cite{FRWCFT1}.



\section{Where Do We Go from Here?}\label{sec_quovadis}

From the preceding discussion we clearly see that we have just begun to scratch the tip of an iceberg's worth of a rich set of models and relationships between various models of large-field inflation in string theory. We are just starting to map the underlying structure of axion monodromy which seems to be a widespread property of the axion sector of string compactifications. Given that large-field inflation with its GUT-scale inflaton potential is maybe our best hope to come close to string-scale physics using observations, we should understand as much as possible about the generic properties of the mechanism while continue to build more explicit sample construction. A synopsis of both, particular constructions and an understanding of the general properties of axion monodromy, might give us what we need to finally have estimates for the inflationary observables $n_s$ and $r$ beyond the level of a few lamp posts.

\section*{Acknowledgements}

I wish to express my deep gratitude to a lot of deep and insightful conversations with many collaborators and friends which helped me most of the way towards the understanding expressed in these notes. This work was supported by the Impuls und Vernetzungsfond of the Helmholtz Association of German Research Centres under grant HZ-NG-603. 

\newpage

\bibliographystyle{ws-rv-van}
\bibliography{stringcosmo}

\printindex                         
\end{document}